\documentclass[conference,letterpaper]{IEEEtran}
\usepackage{fancyhdr,graphicx}
\begin{document}

\title{Microcontroller based distributed and networked control system for public cluster}

\author{\authorblockN{I. Firmansyah, Z. Akbar and L.T. Handoko}
\authorblockA{Group for Theoretical and Computational Physics, Research Center for Physics, Indonesian Institute of Sciences (LIPI)\\
Kompleks Puspiptek Serpong, Tangerang 15310, Indonesia\\
\textit{firmansyah@teori.fisika.lipi.go.id, zaenal@teori.fisika.lipi.go.id, handoko@teori.fisika.lipi.go.id}\\
\textit{http://teori.fisika.lipi.go.id}}}

\maketitle

\thispagestyle{fancy}
\fancyhead{}
\lhead{}
\lfoot{978--1--4244--2287--6/08/\$25.00~\copyright~2008 IEEE}
\cfoot{}
\rfoot{}
\renewcommand{\headrulewidth}{0pt}
\renewcommand{\footrulewidth}{0pt}

\begin{abstract}
We present the architecture and application of the distributed control in public cluster, a parallel machine which is open for public access. Following the nature of public cluster, the integrated distributed control system is fully accessible through network using a user-friendly web interface. The system is intended mainly to control the power of each node in a block of parallel computers provided to certain users. This  is especially important to extend the life-time of related hardwares, and to reduce the whole running and maintainance costs. The system consists of two parts : the master- and node-controllers, and both are connected each other through RS-485 interface. Each node-controller is assigned with a unique address to  distinguish each of them. We also discuss briefly the implementation of the system at the LIPI Public Cluster.
\end{abstract}


\IEEEpeerreviewmaketitle

\section{Introduction}

\begin{figure*}[t]
 \centering
 \includegraphics[width=16cm]{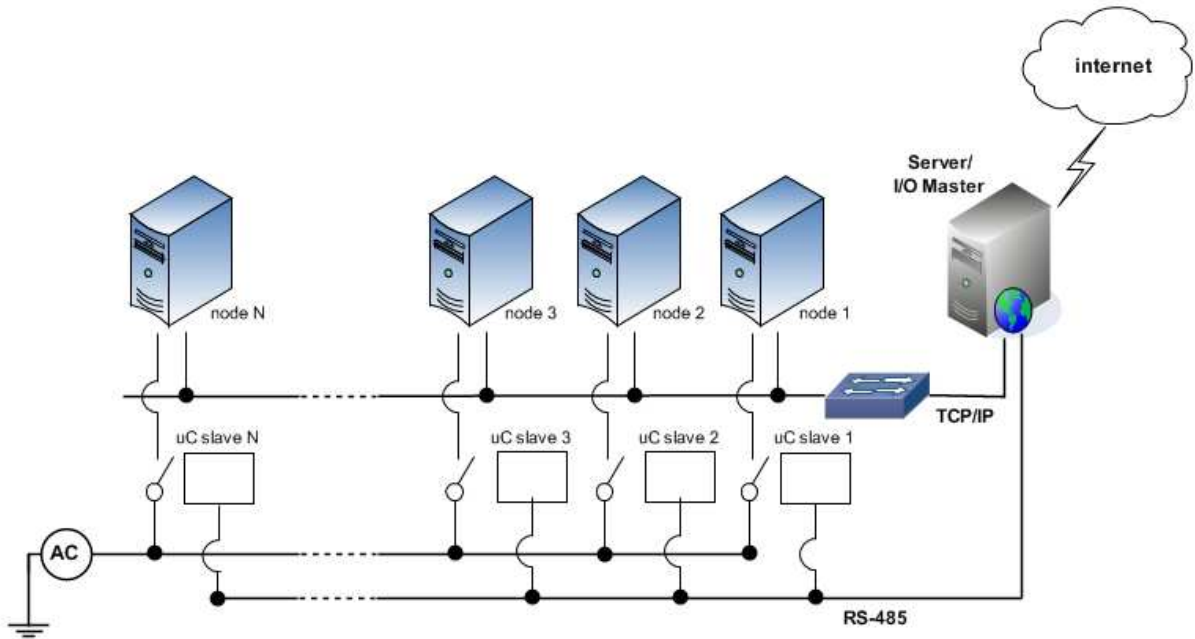}
 \caption{The architecture of the distributed control system at LPC.}
 \label{fig:arsitektur}
\end{figure*}

LIPI Public Cluster (LPC) is a cluster-based computing facility maintained and owned by the Indonesian Institute of Sciences (Lembaga Ilmu Pengetahuan Indonesia - LIPI) \cite{lpc1}. LPC is already running for full operation since 2007 \cite{lpc2}. Although it is still a small scale cluster in the sense of number of nodes already installed, it has unique characteristics among existing clusters around the globe due to its openness. Here ”open” means everyone can access and use it anonymously for free to execute any types of parallel programming. 

In general a cluster is designed to perform a single (huge) computational task at certain period. This makes the cluster system is usually exclusive and not at the level of appropriate cost for most potential users, neither young beginners nor small research groups, especially in developing countries like Indonesia. It is clear that the cluster is in that sense still costly; although there are certain needs to perform such advanced computings. No need to say about educating young generations to be the future users familiar with parallel programming. This background motivates us to further develop an open and free cluster environment for public \cite{lpc3}. 

According to its nature, LPC is in contrast with any conventional clusters, designed to accommodate multiple users with their own parallel programming executed independently at the same period. Therefore an issue on resource allocation is crucial, not only in the sense of allocating hardware to the appropriate users but also to prevent any interference among them \cite{lpc4}.

Concerning its main objective as a training field to learn parallel programming, the public cluster should be accessible and user-friendly for all users with various level of knowledge on parallel programming. It should also have enough flexibility regarding various ways of accessing the system in any platforms as well. This can be achieved by deploying a web-based interface integrating all aspects \cite{lpc6}. Presently we have resolved some issues, such as security from anonymous users to prevent any kinds of interference among different tasks running simultaneously on multi blocks \cite{lpc4}, algorithm for resource allocation management and the distributed control over web for both administrators and end-users \cite{lpc5}.

In term of the hardware, the LPC consists of several PC connected each other through high performance TCP/IP network. Further issue is how to keep the whole performance of cluster during operation especially to control the power supply for each node, such that only selected nodes are activated and the rest nodes are in off-condition. This approach can be accomplished by using the distributed control system. Actually, we have so far deployed more straightforward solution using the centralized microcontroller-based control system  \cite{kontrol}. In this approach, the node is fully controlled by a single control system connected to the power system of each node. This solution is efficient and easy-to-maintain. On the other hand, it leads to difficulty when the control system itself needs regular maintainance, that is we must totally shut down the whole cluster. 

Since the public cluster is running over the time, then it is difficult to set an appropriate down-period. Obviously the present control system requires significant improvements. This is the main reason we are considering the distributed control system \cite{csang}. This requirement is getting more important as now the LPC is enabling connection between its blocks with global grids \cite{grid}.

The paper is organized as follows. First, after this brief introduction we describe the arhitecture of distributed control system under consideration in Sec. \ref{sec:arsitektur}. It is then followed with the implementation in \ref{sec:implementasi} and the communication protocol in our approach in Sec. \ref{sec:komunikasi}. Finally we summarize the paper and provide some future issues and further development.

\section{Architecture}
\label{sec:arsitektur}

\begin{figure*}[t]
 \centering
 \includegraphics[width=16cm]{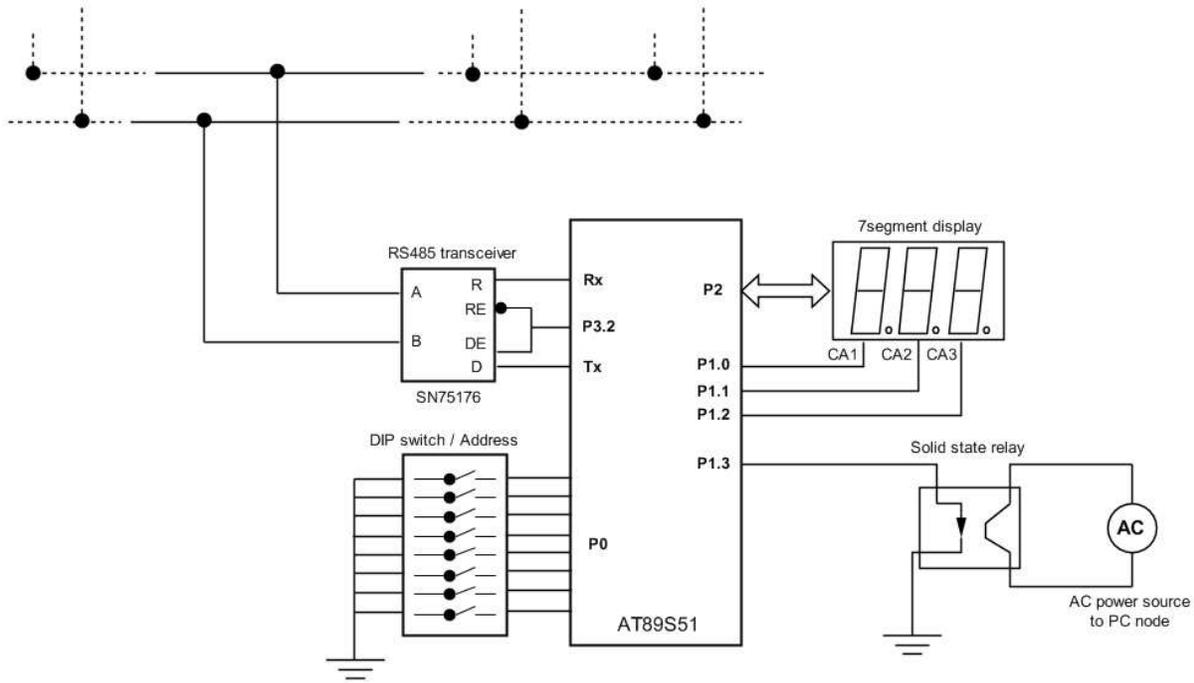}
 \caption{The slave microcontroller circuit for a node-controller.}
 \label{fig:mikro}
\end{figure*}

Concerning the problem mentioned before, we consider the distributed control system with a master-controller and the slaves for node-controllers in each node of public cluster with the following features :
\begin{enumerate}
\item Each node-controller is able to physically shut down each node completely. 
 \item The node-controller for each node is independent each other, that means installing and removing the control system would not interupt another ones in any means.
\item The whole system is controllable remotely through internet.
\end{enumerate}

Therefore, we consider an architecture where the node-controller is attached directly to the power supply in each node independently, while the main one is attached to the master or I/O server. The schematic diagram is given in Fig. \ref{fig:arsitektur}. Subsequently all node-controllers are connected paralelly to the main controller. This approach allows more independent architecture for each sub-control system. 

The distributed control system usually consists of master and several slaves which are connected each other through communication link such as TCP/IP, RS-232, RS-422 and RS-485. The master unit controls the flow of the communication process; and in contrast the slave unit is just receiving any incoming commands from master unit and executing an appropriate action. In many applications, sometimes each slave can be a master and vice versa. However the most important thing is we must deploy the communication protocol with a capability to avoid data collision in the middle of receiving and transmitting the data among them. This communication protocol is very crucial to ensure the master is able to send or request the data or from desired slaves successfully. 

In this paper, we are focused on implementing the distributed control system for LPC. As can be seen in Fig. \ref{fig:arsitektur}, the master controller is attached to the I/O server, while the slaves are to the nodes. Since the I/O server is equipped with complete operating system and communication environments as TCP/IP based networking and web servers, the control system is accessible through internet, or in particular web by public. This would fullfil the last requirement above.

\section{Implementation}
\label{sec:implementasi}

Since both master server and microcontroller are equipped with serial UART, we can use  serial port through UART for communication media between them. For communication media between master- and node-controllers, it is better to use the RS-485 bus as network interface that is suitable for transmitting and receiving data through serial port over long distance connection in noisy environment \cite{mcmillan}.

The slaves consist of microcontroller which has the responsibility for controlling the  power supply in each node using solid state relay. Due to the simple task in each node-controller, the 8-bit AT89S51 microcontroller is used as slave to reduce the cost even though the faster microcontroller \cite{mikro,bhargav}, RISC microcontroller type might be better. Serial RS-485 interface is achieved by SN75176 transceiver from Texas Instrument so that each slave can receive the command sent by master through RS-485 bus \cite{bolic}.

Of course, the commands sent by master on RS-485 bus will not always be directed to a desired slave, but will go to all active slaves at the same bus. Therefore we should define a unique address for each slave microcontroller to distinguish the incoming commands broadcasted to all nodes. There are two alternatives for this purpose, that is : 
\begin{itemize}
 \item inserting the address permanently into node-microcontroller’s ROM, or 
\item using DIP switch as input address for each of them.
\end{itemize}
We have choosen the later option in our case. The reason is, DIP switch is more flexible to especially change the addresses. Because, namely at LPC, the number of active nodes is varying time by time according to the avalaibility.

At LPC, a 3-digit common anode seven segment LED is used in each node-controller as a display interface to make it visually recognizable by administrators. More detail on the slave’s circuit can be seen in Fig. \ref{fig:mikro}.

\section{Communication protocol}
\label{sec:komunikasi}

\begin{figure}[b]
 \centering
 \includegraphics[width=6cm]{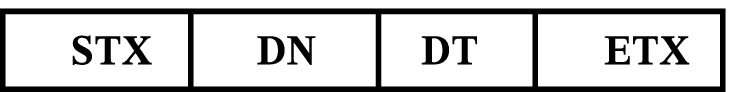}
 \caption{The format of communication protocol.}
 \label{fig:protokol}
\end{figure}

Data transfer among master and several slaves are realized by certain communication protocol. The protocol is designed in such a way that  is sufficient for the master to send the data to certain slave \cite{bolic,vinogradov}. In the current distributed control, the master server executes a program which subsequently send a flagged command, for instance to turn on or off the power of a node, through serial port to the master-controller. The master-controller then redirects the flagged command to all node-controllers. The flag in a command indicates the address of certain node-controller. Although all node-controllers receive the command, only the node-controller with the address fits the flag will execute it, namely turn the node on or off through solid state relay. 

Through the above-mentioned illustration, the flagged messages or command sent by master server should correctly be encoded to allow the appropriate node-controller to translate and execute it. While the rest of node-controllers will do nothing.

The address of each node-controller is assigned using the DIP switch attached on the microcontroller input pins. The address is stored inside the internal memory and is read whenever the microcontroller receives the messages. Changing the DIP switch positions or values will automatically change the address.

Each message or command begins with sync or start bit which contains the address of desired node-controller \cite{csang}. In Fig. \ref{fig:protokol} we show an example of the format of communication protocol. The first byte is Start of Transmission (STX), followed by a slave or device number (DN) that specifies the slave of interest. The next byte is Data Type to Tell (DT) that is the assigned address of the desired slave. Finally the message is terminated with End of Transmission (ETX).

We should note here that the above communication protocol is actually not required for the task presented in this paper, \textit{i.e.} turning the nodes on or off. However, we have indeed deployed the protocol to anticipate a monitoring job through node-controllers. There will be data acquisition processes to retrieve some relevant physical observables like actual external temperature and humidities in each casing.

\section{Summary}

We have discussed how the distributed control system is applied at the LPC, and its architecture. The system has proven a satisfactory performance regarding its original purposes and to improve the up-time of LPC. We have succeeded in implementing a totally independent control system for each node which is very crucial in any cluster machine. However, the system can be realized at low cost, not only at the initial development but also for further maintainance.

The whole system is currently running well at LPC, and has been integrated in a user-friendly web-interface to enable all LPC's users a full access to their allocated blocks of parallel machines \cite{lpc6}.

Now the system is expanded to not only control the power, but also to retrieve and monitor the neighbouring physical observables like temperature and so forth. However for more advanced applications like monitoring system where the speed of data transfer is getting crucial, an RISC type microcontroller like AVR might be more suitable. This work is still in progress.

\section*{Acknowledgment}

The work is partially supported by the Riset Kompetitif LIPI in fiscal year 2009.

\end{document}